# Foundations of a live data exploration environment

## Tomas Petricek[a,b]

a   University of Kent, UK
b   and The Alan Turing Institute, UK

**Abstract**   A growing amount of code is written to explore and analyze data, often by data analysts who do not have a traditional background in programming, for example by journalists. The way such data anlysts write code is different from the way software engineers do so. They use few abstractions, work interactively and rely heavily on external libraries. We aim to capture this way of working and build a programming environment that makes data exploration easier by providing instant live feedback.

We combine theoretical and applied approach. We present the *data exploration calculus*, a formal language that captures the structure of code written by data analysts. We then implement a data exploration environment that evaluates code instantly during editing and shows previews of the results. We formally describe an algorithm for providing instant previews for the data exploration calculus that allows the user to modify code in an unrestricted way in a text editor. Supporting interactive editing is tricky as any edit can change the structure of code and fully recomputing the output would be too expensive. We prove that our algorithm is correct and that it reuses previous results when updating previews after a number of common code edit operations. We also illustrate the practicality of our approach with an empirical evaluation and a case study.

As data analysis becomes an ever more important use of programming, research on programming languages and tools needs to consider new kinds of programming workflows appropriate for those domains and conceive new kinds of tools that can support them. The present paper is one step in this important direction.

## The Art, Science, and Engineering of Programming

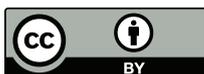

## 1   Introduction

One of the aspects that make spreadsheets easier to use than other programming tools is their liveness. When you change a cell in Excel, the whole spreadsheet updates instantly and you immediately see new results, without having to explicitly trigger re-computation and without having to wait for an extensive period of time.

An increasing number of programming environments aim to provide a live development experience for standard programming languages, but doing this is not easy. Fully recomputing the whole program after every keystroke is inefficient and calculating how a change in the source code changes the result is extremely hard when the text editor allows arbitrary changes. Consider the following snippet that gets the release years of the 10 most expensive movies from a data set movies:

```
let top = movies.sortBy(λx → x.getBudget())
    .take(10).map(λx → x.getReleased().format("yyyy"))
```

A live programming environment computes and displays the list of years. Suppose that the programmer then makes 10 a variable and changes the date format:

```
let count = 10
let top = movies.sortBy(λx → x.getBudget())
        .take(count).map(λx → x.getReleased().format("dd-mm-yyyy"))
```

Ideally, the live programming environment should understand the change, reuse a cached result of the first two transformations (sorting and taking the first 10 elements) and only evaluate the map operation to differently format the release dates of top 10 movies. Our environment does this for a simple data exploration language in an unrestricted text editor. We discuss related work in section 7, but we briefly review the most important directions here, in order to situate our contributions.

### Contributions

We present the design and implementation of a live programming environment for a simple data exploration language that provides correct and efficient instant feedback, yet is integrated into an ordinary text editor. Our key contributions are:

- We introduce the *data exploration calculus* (section 3), a small formally tractable language for data exploration. The calculus is motivated by our review of how data analysts work (section 2) and makes it possible to write simple, transparent and reproducible data analyses.

- A live programming environment does not operate in batch mode and so we cannot follow classic compiler literature. We capture the essence of such new perspective (section 4) and use it to build our *instant preview* mechanism (section 5) that evaluates code instantly during editing.

- We prove that our instant preview mechanism is correct (section 6.1) and that it reuses previously evaluated values when possible (section 6.2). We illustrate the practicality of the mechanism using an empirical evaluation (section 6.3) and a case study (section 6.4).

## UN Comtrade exports data

```python
material = 'plastics' # 'plastics', 'paper'
```

## Loading exports data

```python
df_mat = pd.read_csv('{material}-2017.csv').fillna(0).sort_values(['country_name', 'period'])
df_mat.head()
```

|   | period | country_name | kg | country_code |
|---|--------|--------------|------|--------------|
| **0** | 2017-01-01 | Algeria | 43346.0 | 12 |
| **1** | 2017-03-01 | Algeria | 32800.0 | 12 |
| **2** | 2017-03-01 | Antigua and Barbuda | 17000.0 | 28 |

## Join to country codes

```python
# Set keep_default_na because the Namibia has ISO code NA
df_isos = pd.read_excel('iso.xlsx', keep_default_na=False).drop_duplicates('country_code')
df = df_mat.copy().merge(df_isos, 'left', 'country_code').rename({ 'iso2': 'country_code' }, axis=1)
df.head()
```

|   | period | country_name | kg | country_code |
|---|--------|--------------|------|--------------|
| **0** | 2017-01-01 | Algeria | 43346.0 | DZ |
| **1** | 2017-03-01 | Algeria | 32800.0 | DZ |
| **2** | 2017-03-01 | Antigua and Barbuda | 17000.0 | AG |

■ **Figure 1**  Financial Times analysis that joins UN trade database with ISO country codes.

### 2   Understanding how data scientists work

Data scientists often use general-purpose programming languages such as Python, but the kind of code they write and the way they interact with the system is very different from how software engineers work [21]. This paper focuses on simple data wrangling and data exploration as done, for example, by journalists analysing government datasets. This new kind of non-expert programmers is worth our attention as they often work on informing the public. They need easy-to-use tools, but not necessarily a full programming langugae. In this section, we ilustrate how such data analyses look and we provide a justification for the design of our data exploration calculus.

#### 2.1  Simple data exploration in Jupyter

Data analysts increasingly use notebook systems such as Jupyter, which make it possible to combine text, equations and code with results of running the code, such as tables or visualizations. Notebooks blur the conventional distinction between development and execution. Data analysts write small snippets of code, run them to see results immediately and then revise them.

Notebooks are used by users ranging from scientists who implement complex models of physical systems to journalists who perform simple data aggregations and create visualizations. Our focus is on the simplest use cases. Making programmatic data exploration more spreadsheet-like should encourage users to choose programming tools over spreadsheets, resulting in more reproducible and transparent data analyses.

Consider the Financial Times analysis of plastic waste [7, 25]. It joins datasets from Eurostat, UN Comtrade and more, aggregates the data and builds a visualization comparing waste flows in 2017 and 2018. Figure 1 shows an excerpt from one notebook of the data analysis. The code has a number of important properties:

- There is no abstraction. The analysis uses lambda functions as arguments to library calls, but it does not define any custom functions. Code is parameterized by having a global variable `material` set to `"plastics"` and keeping other possible values in a comment. This lets the analyst run and check results of intermediate steps.

- The code relies on external libraries. Our example uses Pandas [36], which provides operations for data wrangling such as `merge` to join datasets or `drop_duplicates` to delete rows with duplicate column values. Such standard libraries are external to the data analysis and are often implemented in another language like C++.

- The code is structured as a sequence of commands. Some commands define a variable, either by loading data, or by transforming data loaded previously. Even in Python, data is often treated as immutable. Other commands produce an output that is displayed in the notebook.

- There are many corner cases, such as the fact that `keep_default_na` needs to be set to handle Namibia correctly. These are discovered interactively by running the code and examining the output, so providing an instant feedback is essential.

Many Jupyter notebooks are more complex than the above example and might use helper functions or object-oriented code. However, simple data analyses such as the one discussed here are frequent enough and pose interesting problems for programming tools. This paper aims to bring such analyses to the attention of programming research community by capturing their essential properties as a formal calculus.

## 2.2 Dot-driven data exploration in The Gamma

Simple data exploration has been the motivation for a scripting language The Gamma [45]. Scripts in The Gamma are sequences of commands that either define a variable or produce an output. It does not support top-level functions and lambda functions can be used only as method arguments. Given the limited expressiveness of The Gamma, libraries are implemented in other languages, such as JavaScript. The Gamma uses type providers [54] for accessing external data sources. Type providers generate object types with members and The Gamma offers those in an auto-complete list when the user types dot ('.') to access a member. The combination of type providers and auto-complete makes it possible to solve a large number of data exploration tasks through the very simple interaction of selecting operations from a list.

**(a)** The analysis counts the number of distinct events per athlete. After typing '.' the editor offers further aggregation operations.

**(b)** Our live programming environment for The Gamma. The table is updated on-the-fly and shows the result at the current cursor position.

■ **Figure 2**  Previous work on The Gamma with auto-complete based on type information (left) and our new editor with instant preview (right).

The example in figure 2a summarizes data on Olympic medals. Identifiers such as 'sum Bronze' are names of members generated by the type provider. The type provider used in this example generates an object with members for data transformations such as 'group data', which return further objects with members for specifying transformation parameters, such as selecting the grouping key using 'by Athlete'.

The Gamma language is richer, but the example in figure 2a shows that non-trivial data exploration can be done using a very simple language. The assumptions about structure of code that are explicit in The Gamma are implicitly present in Python and R data analyses produced by journalists, economists and other users with other than programming background. When we refer to The Gamma in this paper, readers not familiar with it can consider a small subset of Python.

### 2.3  Live programming for data exploration

The implementation that accompanies this paper builds a live programming environment for The Gamma. It is discussed in section 6.4 and it replaces the original text editor with just auto-complete with a live programming environment that provides *instant previews*.

An example of a instant preview is shown in figure 2b. As noted earlier, The Gamma programs consist of lists of commands which are either expressions or let bindings. Our editor displays a instant preview below the command that the user is currently editing. The preview shows the result of evaluating the expression or the value assigned to a bound variable. When the user changes the code, the preview is updated automatically, without any additional interaction with the user.

## Foundations of a live data exploration environment

**Programs, commands, terms, expressions and values**

$$p ::= c_1; \ldots; c_n \qquad t ::= o \mid x \qquad e ::= t \mid \lambda x \to e$$
$$c ::= \mathsf{let}\ x = t \mid t \qquad \mid \ t.m(e, \ldots, e) \qquad v ::= o \mid \lambda x \to e$$

**Evaluation contexts of expressions**

$$C_e[-] = C_e[-].m(e_1, \ldots, e_n) \mid o.m(v_1, \ldots, v_m, C_e[-], e_1, \ldots, e_n) \mid -$$
$$C_c[-] = \mathsf{let}\ x = C_e[-] \mid C_e[-]$$
$$C_p[-] = o_1; \ldots; o_k; C_c[-]; c_1; \ldots; c_n$$

**Let elimination and member reduction**

$$o_1; \ldots; o_k; \mathsf{let}\ x = o; c_1; \ldots; c_n \rightsquigarrow$$
$$o_1; \ldots; o_k; o; c_1[x \leftarrow o]; \ldots; c_n[x \leftarrow o] \qquad \text{(let)}$$

$$\frac{o.m(v_1, \ldots, v_n) \rightsquigarrow_\epsilon o'}{C_p[o.m(v_1, \ldots, v_n)] \rightsquigarrow C_p[o']} \qquad \text{(external)}$$

■ **Figure 3** Syntax, contexts and reduction rules of the data exploration calculus

There are a number of guiding principles that inform our design. First, we allow the analyst to edit code in an unrestricted form in a text editor. Although structured editors provide an appealing alternative and make recomputation easier, we prefer the flexibility of plain text. Second, we focus on the scenario when code changes, but input does not. Rapid feedback allows the analyst to quickly adapt code to correctly handle corner cases that typical analysis involves. In contrast to work on incremental computation, we do not consider the case when data changes, although supporting interactive data exploration of streaming data is an interesting future direction.

## 3 Data exploration calculus

The *data exploration calculus* is a small formal language for data exploration. The calculus is not, in itself, Turing-complete and it can only be used together with external libraries that define what objects are available and what the behaviour of their members is. This is sufficient to capture the simple data analyses discussed in section 2. We define the calculus in this section and then use it to formalise our instant preview mechanism in section 4. The instant preview mechanism does not rely on types and so we postpone the discussion of static typing to appendix C. Interestingly, it reuses the mechanism used for live previews.

### 3.1 Language syntax

The calculus combines object-oriented features such as member access with functional features including lambda functions. The syntax is defined in figure 3. Object values $o$ are defined by external libraries that are used in conjunction with the core calculus.

A program $p$ in the data exploration calculus consists of a sequence of commands $c$. A command can be either a let binding or a term. Let bindings define variables $x$ that can be used in subsequent commands. Lambda functions can only appear as arguments in method calls. A term $t$ can be a value, variable or a member access, while an expression $e$, which can appear as an argument in member access, can be a lambda function or a term.

### 3.2 Operational semantics

The data exploration calculus is a call-by-value language. We model evaluation as a small-step reduction $\rightsquigarrow$. Fully evaluating a program results in an irreducible sequence of objects $o_1; \ldots; o_n$ (one object for each command, including let bindings) which can be displayed as intermediate results of the data analysis. The operational semantics is parameterized by a relation $\rightsquigarrow_\epsilon$ that models the functionality of the external libraries used with the calculus and defines the reduction behaviour for member accesses. The relation has the following form:

$$o_1.m(v_1, \ldots, v_n) \rightsquigarrow_\epsilon o_2$$

Here, the operation $m$ is invoked on an object and takes values (objects or function values) as arguments. The reduction always results in an object. Figure 3 defines the reduction rules in terms of $\rightsquigarrow_\epsilon$ and evaluation contexts; $C_e$ specifies left-to-right evaluation of arguments of a method call, $C_c$ specifies evaluation of a command and $C_p$ defines left-to-right evaluation of a program. The rule (external) calls a method provided by an external library in a call-by-value fashion; (let) substitutes a value of an evaluated variable in all subsequent commands and leaves the result in the list of commands. Note that our semantics does not define how $\lambda$ applications are reduced. This is done by external libraries, which will typically supply functions with arguments using standard $\beta$-reduction. The behaviour is subject to constraints discussed next.

### 3.3 Example external library

The data exploration calculus is not limited to the data exploration domain. It can be used with external libraries for a wide range of other simple programming tasks, such as image manipulation, as done in section 6.3. However, we choose a name that reflects the domain that motivated this paper. To illustrate how a definition of an external library looks, consider the following simple data manipulation script:

```
let l = list.range(0, 10)
l.map(λx → math.mul(x, 10))
```

An external library provides the list and math values with members range, map and mul. The objects of the external library consist of numbers $n$, lists of objects $[o_1, \ldots, o_k]$ and failed computations $\perp$ [37]. Next, the external library needs to define the $\rightsquigarrow_\epsilon$ relation that defines the evaluation of member accesses. The following shows the rules for members of lists, assuming the only supported member is map:

$$\frac{e[x \leftarrow n_i] \rightsquigarrow o_i \quad (\textit{for all } i \in 1 \dots k)}{[n_1, \dots, n_k].\mathsf{map}(\lambda x \rightarrow e) \rightsquigarrow_\epsilon [o_1, \dots, o_k]} \qquad \frac{(\textit{otherwise})}{[n_1, \dots, n_k].m(v_1, \dots, v_n) \rightsquigarrow_\epsilon \perp}$$

When evaluating map, we apply the provided function to all elements of the list using standard $\beta$-reduction, defined recursively using $\rightsquigarrow$, and return a list with resulting objects. The $\rightsquigarrow_\epsilon$ relation is defined on all member accesses, but non-existent members reduce to the failed computation $\perp$.

### 3.4 Properties

The data exploration calculus has a number of desirable properties. Some of those require that the relation $\rightsquigarrow_\epsilon$, which defines evaluation for external libraries, satisfies a number of conditions. We discuss *normalization* and *let elimination* in this section. Those two are particularly important as they will allow us to prove correctness of our method of evaluating instant previews in section 6.1.

**Definition 1** (Further reductions). We define two additional reduction relations:

- We write $\rightsquigarrow^*$ for the reflexive, transitive closure of $\rightsquigarrow$
- We write $\rightsquigarrow_{\mathsf{let}}$ for a call-by-name let binding elimination $c_1; \dots; c_{k-1};$ let $x = t; c_{k+1}; \dots; c_n \rightsquigarrow_{\mathsf{let}} c_1; \dots; c_{k-1}; t; c_{k+1}[x \leftarrow t]; \dots; c_n[x \leftarrow t]$

We say that two expressions $e$ and $e'$ are *observationally equivalent* if, for any context $C$, the expressions $C[e]$ and $C[e']$ reduce to the same value. Lambda functions $\lambda x \rightarrow 2$ and $\lambda x \rightarrow 1+1$ are not equal, but they are observationally equivalent. We require that external libraries satisfy two conditions. First, when a method is called with observationally equivalent values as arguments, it should return the same value. Second, the evaluation of $o.m(v_1, \dots, v_n)$ should be defined for all $o, n$ and $v_i$. The definition in section 3.3 satisfies those by using standard $\beta$-reduction for lambda functions and by reducing all invalid calls to the $\perp$ object.

**Definition 2** (External library). An external library consists of a set of objects $O$ and a reduction relation $\rightsquigarrow_\epsilon$ that satisfies the following two properties:

**Totality** For all $o, m, i$ and all $v_1, \dots, v_i$, there exists $o'$ such that $o.m(v_1, \dots, v_i) \rightsquigarrow_\epsilon o'$.

**Compositionality** For observationally equivalent arguments, the reduction should always return the same object, i.e. given $e_0, e_1, \dots, e_n$ and $e'_0, e'_1, \dots, e'_n$ and $m$ such that $e_0.m(e_1, \dots, e_n) \rightsquigarrow^* o$ and $e'_0.m(e'_1, \dots, e'_n) \rightsquigarrow^* o'$ then if for any contexts $C_0, C_1, \dots, C_n$ it holds that if $C_i[e_i] \rightsquigarrow^* o_i$ and $C_i[e'_i] \rightsquigarrow^* o_i$ for some $o_i$ then $o = o'$.

Compositionality is essential for proving the correctness of our instant preview mechanism and implies determinism of external libraries. Totality allows us to prove normalization, i.e. all programs reduce to a value – although the resulting value may be an error value provided by the external library.

**Theorem 1** (Normalization). *For all $p$, there exists $n, o_1, \ldots, o_n$ such that $p \rightsquigarrow^* o_1; \ldots; o_n$.*

*Proof.* A program that is not a sequence of values can be reduced and reduction decreases the size of the program. See appendix A.1 for more detail. □

Although the reduction rules (let) and (external) of the data exploration calculus define an evaluation in a call-by-value order, eliminating let bindings in a call-by-name way using the $\rightsquigarrow_{\text{let}}$ reduction does not affect the result. This simplifies our later proof of instant preview correctness in section 6.1.

**Lemma 2** (Let elimination for a program). *Given any program $p$ such that $p \rightsquigarrow^* o_1; \ldots; o_n$ for some $n$ and $o_1, \ldots, o_n$ then if $p \rightsquigarrow_{\text{let}} p'$ for some $p'$ then also $p' \rightsquigarrow^* o_1; \ldots; o_n$.*

*Proof.* By constructing $p' \rightsquigarrow^* o_1; \ldots; o_n$ from $p \rightsquigarrow^* o_1; \ldots; o_n$. See appendix A.2. □

## 4 Formalising a live programming environment

A naive way of providing instant previews during code editing would be to re-evaluate the code after each change. This would be wasteful – when writing code to explore data, most changes are additive. To update a preview, we only need to evaluate newly added code. We describe an efficient mechanism in this section.

### 4.1 Maintaining the dependency graph

The key idea behind our method is to maintain a dependency graph [32] with nodes representing individual operations of the computation that can be evaluated to obtain a preview. Each time the program text is modified, we parse it afresh (using an error-recovering parser) and bind the abstract syntax tree to the dependency graph. When binding a new expression to the graph, we reuse previously created nodes as long as they have the same structure and the same dependencies. For expressions that have a new structure, we create new nodes.

The nodes of the graph serve as unique keys into a lookup table containing previously evaluated parts of the program. When a preview is requested for an expression, we use the graph node bound to the expression to find a preview. If a preview has not been evaluated, we force the evaluation of all dependencies in the graph and then evaluate the operation represented by the current node.

#### 4.1.1 Elements of the graph

The nodes of the graph represent individual operations of the computation. In our design, the nodes are used as cache keys, so we attach a unique symbol $s$ to some of the nodes. That way, we can create two unique nodes representing, for example, access to a member named take which differ in their dependencies.

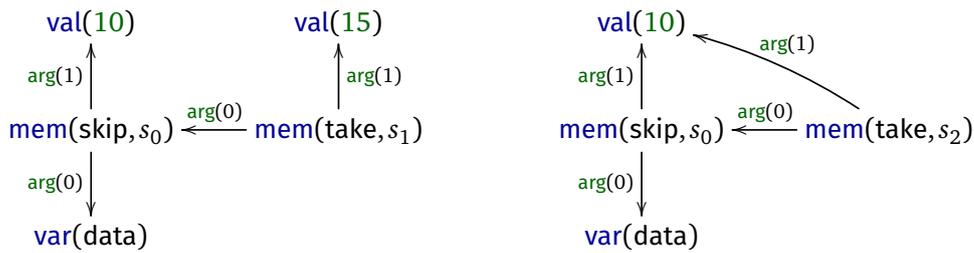

**(a)** Graph constructed from initial expression:
    let $x$ = 15 in data.skip(10).take($x$)

**(b)** Updated graph after changing $x$ to 10:
    let $x$ = 10 in data.skip(10).take($x$)

■ **Figure 4**   Dependency graphs formed by two steps of the live programming process.

The graph edges are labelled with labels indicating the kind of dependency. For a method call, the labels are "first argument", "second argument" and so on. Writing $s$ for symbols and $i$ for integers, nodes (vertices) $v$ and edge labels $l$ are defined as:

$$\begin{aligned} v &= \text{val}(o) \mid \text{var}(x) \mid \text{mem}(m,s) \mid \text{fun}(x,s) & \text{(\textit{Vertices})} \\ l &= \text{body} \mid \text{arg}(i) & \text{(\textit{Edge labels})} \end{aligned}$$

The val node represents a primitive value and contains the object itself. Two occurrences of 10 in the source code will be represented by the same node. Member access mem contains the member name, together with a unique symbol $s$ – two member access nodes with different dependencies will contain different symbol. Dependencies of member access are labelled with arg indicating the index of the argument (0 for the instance and 1, . . . for the arguments). Finally, nodes fun and var represent function values and variables bound by $\lambda$ abstraction.

### 4.1.2 Example graph

figure 4 illustrates how we build the dependency graph. Node representing take($x$) depends on the argument – the number 15 – and the instance, which is a node representing skip(10). This depends on the instance data and the number 10. Note that variables bound via let binding such as $x$ do not appear as var nodes. The node using it depends directly on the node representing the expression assigned to $x$.

After changing the value of $x$, we create a new graph. The dependencies of the node mem(skip, $s_0$) are unchanged and so the symbol $s_0$ attached to the node remains the same and previously computed previews can be reused. This part of the program is not recomputed. The arg(1) dependency of the take call changed and so we create a new node mem(skip, $s_2$) with a fresh symbol $s_2$. The preview for this node is then computed as needed using the already known values of its dependencies.

### 4.1.3 Reusing graph nodes

The binding process takes an expression and constructs a dependency graph. It uses a lookup table to reuse previously created member access and function value nodes. The key in the lookup table is formed by a node kind together with a list of dependencies.

$$\text{bind-expr}_{\Gamma,\Delta}(e_0.m(e_1,\ldots,e_n)) = v, (\{v\} \cup V_0 \cup \ldots \cup V_n, E \cup E_0 \cup \ldots \cup E_n) \qquad (1)$$

when $v_i, (V_i, E_i) = \text{bind-expr}_{\Gamma,\Delta}(e_i)$ and $v = \Delta(\text{mem}(m), [(v_0, \text{arg}(0)), \ldots, (v_n, \text{arg}(n))])$

let $E = \{(v, v_0, \text{arg}(0)), \ldots, (v, v_n, \text{arg}(n))\}$

$$\text{bind-expr}_{\Gamma,\Delta}(e_0.m(e_1,\ldots,e_n)) = v, (\{v\} \cup V_0 \cup \ldots \cup V_n, E \cup E_0 \cup \ldots \cup E_n) \qquad (2)$$

when $v_i, (V_i, E_i) = \text{bind-expr}_{\Gamma,\Delta}(e_i)$ and $\Delta(\text{mem}(m), [(v_0, \text{arg}(0)), \ldots, (v_n, \text{arg}(n))]) \downarrow$

let $v = \text{mem}(m, s), s$ fresh and $E = \{(v, v_0, \text{arg}(0)), \ldots, (v, v_n, \text{arg}(n))\}$

$$\text{bind-expr}_{\Gamma,\Delta}(\lambda x \to e) = v, (\{v\} \cup V, \{e\} \cup E) \qquad (3)$$

when $\Gamma_1 = \Gamma \cup \{x, \text{var}(x)\}$ and $v_0, (V, E) = \text{bind-expr}_{\Gamma_1,\Delta}(e)$ and $v = \Delta(\text{fun}(x), [(v_0, \text{body})])$

let $e = (v, v_0, \text{body})$

$$\text{bind-expr}_{\Gamma,\Delta}(\lambda x \to e) = v, (\{v\} \cup V, \{e\} \cup E) \qquad (4)$$

when $\Gamma_1 = \Gamma \cup \{x, \text{var}(x)\}$ and $v_0, (V, E) = \text{bind-expr}_{\Gamma_1,\Delta}(e)$ and $\Delta(\text{fun}(x), [(v_0, \text{body})]) \downarrow$

let $v = \text{fun}(x, s), s$ fresh and $e = (v, v_0, \text{body})$

$$\text{bind-expr}_{\Gamma,\Delta}(o) = \text{val}(o), (\{\text{val}(o)\}, \emptyset) \qquad (5)$$

$$\text{bind-expr}_{\Gamma,\Delta}(x) = v, (\{v\}, \emptyset) \text{ when } v = \Gamma(x) \qquad (6)$$

■ **Figure 5**  Binding rules that define a construction of a dependency graph for an expression.

A node kind includes the member or variable name; a lookup table $\Delta$ then maps a node kind with a list of dependencies to a cached node:

$$k ::= \text{fun}(x) \mid \text{mem}(m) \qquad \textit{(Node kinds)}$$
$$\Delta(k, [(v_1, l_1), \ldots, (v_n, l_n)]) \qquad \textit{(Lookup for a node)}$$

The example on the second line looks for a node of a kind $k$ that has dependencies $v_1, \ldots, v_n$ labelled with labels $l_1, \ldots, l_n$. We write $\Delta(k, l) \downarrow$ when a value for a key is not defined. When creating the graph in figure 4b, we perform the following lookups:

$$\Delta(\text{mem}(\text{skip}), [(\text{var}(\text{data}), \text{arg}(0)), (\text{val}(10), \text{arg}(1))]) \qquad (1)$$
$$\Delta(\text{mem}(\text{take}), [(\text{mem}(\text{skip}, s_0), \text{arg}(0)), (\text{val}(10), \text{arg}(1))]) \qquad (2)$$

First, we look for the skip member access. The result is the $\text{mem}(\text{skip}, s_0)$ known from the previous step. We then look for the take member access. In the earlier step, the argument of take was 15 rather than 10 and so this lookup fails. We then construct a new node $\text{mem}(\text{take}, s_2)$ and later add it to the cache.

## 4.2 Binding expressions to a graph

After parsing modified code, we update the dependency graph and link each node of the abstract syntax tree to a node of the dependency graph. This process is called binding and is defined by the bind-expr function (figure 5) and bind-prog function (figure 6). Both functions are annotated with a lookup table $\Delta$ and a variable context $\Gamma$.

$$\text{bind-prog}_{\Gamma,\Delta}(\text{let } x = e; c_2; \ldots; c_n) = v_1; \ldots; v_n, (\{v_1\} \cup V \cup V_1, E \cup E_1) \tag{7}$$
$$\quad \text{let } v_1, (V_1, E_1) = \text{bind-expr}_{\Gamma,\Delta}(e_1) \text{ and } \Gamma_1 = \Gamma \cup \{(x, v_1)\}$$
$$\quad \text{and } v_2; \ldots; v_n, (V, E) = \text{bind-prog}_{\Gamma_1,\Delta}(c_2; \ldots; c_n)$$
$$\text{bind-prog}_{\Gamma,\Delta}(e; c_2; \ldots; c_n) = v_1; \ldots; v_n, (\{v_1\} \cup V \cup V_1, E \cup E_1) \tag{8}$$
$$\quad \text{let } v_1, (V_1, E_1) = \text{bind-expr}_{\Gamma,\Delta}(e) \text{ and } v_2; \ldots; v_n, (V, E) = \text{bind-prog}_{\Gamma_1,\Delta}(c_2; \ldots; c_n)$$
$$\text{bind-prog}_{\Gamma,\Delta}([]) = [], (\emptyset, \emptyset) \tag{9}$$

■ **Figure 6**  Binding rules that define a construction of a dependency graph for a program.

$$\text{update}_{V,E}(\Delta_{i-1}) = \Delta_i \text{ such that:}$$
$$\quad \Delta_i(\text{mem}(m), [(v_0, \text{arg}(0)), \ldots, (v_n, \text{arg}(n))]) = \text{mem}(m, s)$$
$$\quad\quad \text{when } \text{mem}(m, s) \in V \text{ and } (\text{mem}(m, s), v_i, \text{arg}(i)) \in E \text{ for } i \in 0, .., n$$
$$\quad \Delta_i(\text{fun}(x), [(v_1, \text{body})]) = \text{fun}(x, s)$$
$$\quad\quad \text{when } \text{fun}(x, s) \in V \text{ and } (\text{fun}(x, s), v_1, \text{body}) \in E$$
$$\quad \Delta_i(v) = \Delta_{i-1}(v) \quad \text{(otherwise)}$$

■ **Figure 7**  Updating the node cache after binding a new graph

The variable context is a map from variable names to dependency graph nodes and is used for variables bound using let binding.

When invoked, $\text{bind-expr}_{\Gamma,\Delta}(e)$ returns a node $v$ that corresponds to the expression $e$, paired with a dependency graph $(V, E)$ formed by nodes $V$ and labelled edges $E$. That edges are written as $(v_1, v_2, l)$ and include a label $l$. The $\text{bind-prog}_{\Gamma,\Delta}$ function works similarly, but turns a sequence of commands into a sequence of nodes.

When binding a member access, we use bind-expr recursively to get a node and a dependency graph for each sub-expression. The nodes representing sub-expressions are then used for lookup into $\Delta$, together with their labels. If a node already exists in $\Delta$ it is reused (1). Alternatively, we create a new node containing a fresh symbol (2). The graph node bound to a function depends on a synthetic node $\text{var}(x)$ that represents a variable of unknown value. When binding a function, we create a variable node and add it to the variable context $\Gamma_1$ before binding the body. As with member access, the node representing a function may (3) or may not (4) already exist in the lookup table.

When binding a program, we bind the first command and recursively process remaining commands (9). For let binding (7), we bind the expression $e$ assigned to the variable to obtain a graph node $v_1$. We then store the node in the variable context $\Gamma_1$ and bind the remaining commands. The variable context is used when binding a variable in bind-expr (6) and so all variables declared using let will be bound to a graph node representing the value assigned to the variable. When the command is just an expression (8), we bind the expression using bind-expr.

### 4.3 Edit and rebind loop

During editing, the dependency graph is repeatedly updated according to the binding rules. We maintain a series of lookup table states $\Delta_0, \Delta_1, \Delta_2, \ldots$ The initial lookup table is empty, i.e. $\Delta_0 = \emptyset$. At a step $i$, we parse a program $p_i$ and obtain a new dependency graph using the previous $\Delta$. The result is a sequence of nodes corresponding to commands of the program and a graph $(V, E)$:

$$v_1; \ldots; v_n, (V, E) = \text{bind-prog}_{\emptyset, \Delta_{i-1}}(p_i)$$

The new state of the cache is computed using $\text{update}_{V,E}(\Delta_{i-1})$ defined in figure 7. The function adds newly created nodes from the graph $(V, E)$ to the previous cache $\Delta_{i-1}$ and returns a new cache $\Delta_i$.

## 5 Computing instant previews

The binding process constructs a dependency graph after code changes. The nodes in the dependency graph correspond to individual operations that will be performed when running the program. When evaluating a preview, we attach partial results to nodes of the graph. Since the binding process reuses nodes, previews for sub-expressions attached to graph nodes will also be reused.

In this section, we describe how previews are evaluated. The evaluation is done over the dependency graph, rather than over the structure of program expressions as in the operational semantics given in section 3.2. In section 6, we prove that resulting previews are the same as the result we would get by directly evaluating code and we also show that no recomputation occurs when code is edited in certain ways.

### 5.1 Previews and delayed previews

Programs in the data exploration calculus consist of sequence of commands. Those are evaluated to a value with a preview that can be displayed to the user. However, we also support previews for sub-expressions. This can be problematic if the current sub-expression is inside the body of a function. For example:

```
let top = movies.take(10).map(λx → x.getReleased().format("dd-mm-yyyy"))
```

Here, we can directly evaluate sub-expressions movies and movies.take(10), but not $x$.getReleased() because it contains a free variable $x$. Our preview evaluation algorithm addresses this by producing two kinds of previews. A *fully evaluated preview* is just a value, while a *delayed preview* is a partially evaluated expression with free variables:

$$
\begin{aligned}
p &= o \mid \lambda x \to e & \text{(Fully evaluated previews)} \\
d &= p \mid [\![e]\!]_\Gamma & \text{(Evaluated and delayed previews)}
\end{aligned}
$$

A fully evaluated preview $p$ can be either a primitive object or a function value with no free variables. A possibly delayed preview $d$ can be either an evaluated preview $p$

$$(\text{lift-expr}) \quad \frac{v \Downarrow [\![e]\!]_\Gamma}{v \Downarrow_{\text{lift}} [\![e]\!]_\Gamma}$$

$$(\text{fun-val}) \quad \frac{(\text{fun}(x,s), v, \text{body}) \in E \quad v \Downarrow p}{\text{fun}(x,s) \Downarrow \lambda x \to p}$$

$$(\text{lift-prev}) \quad \frac{v \Downarrow p}{v \Downarrow_{\text{lift}} [\![p]\!]_\emptyset}$$

$$(\text{fun-bind}) \quad \frac{(\text{fun}(x,s), v, \text{body}) \in E \quad v \Downarrow [\![e]\!]_x}{\text{fun}(x,s) \Downarrow \lambda x \to e}$$

$$(\text{val}) \quad \frac{}{\text{val}(o) \Downarrow o}$$

$$(\text{fun-expr}) \quad \frac{(\text{fun}(x,s), v, \text{body}) \in E \quad v \Downarrow [\![e]\!]_{x,\Gamma}}{\text{fun}(x,s) \Downarrow [\![\lambda x \to e]\!]_\Gamma}$$

$$(\text{var}) \quad \frac{}{\text{var}(x) \Downarrow [\![x]\!]_x}$$

$$(\text{mem-val}) \quad \frac{\forall i \in \{0 \ldots k\}.(\text{mem}(m,s), v_i, \text{arg}(i)) \in E \quad v_i \Downarrow p_i \quad p_0.m(p_1, \ldots, p_k) \rightsquigarrow_\epsilon p}{\text{mem}(m,s) \Downarrow p}$$

$$(\text{mem-expr}) \quad \frac{\forall i \in \{0 \ldots k\}.(\text{mem}(m,s), v_i, \text{arg}(i)) \in E \quad \exists j \in \{0 \ldots k\}.v_j \not\Downarrow p_j \quad v_i \Downarrow_{\text{lift}} [\![e_i]\!]_{\Gamma_i}}{\text{mem}(m,s) \Downarrow [\![e_0.m(e_1, \ldots, e_k)]\!]_{\Gamma_0, \ldots, \Gamma k}}$$

■ **Figure 8**  Rules that define evaluation of previews over a dependency graph for a program or an expression $e$ that requires variables $\Gamma$. We use an untyped language and so $\Gamma$ is just a list of variables $x_1, \ldots, x_n$. As discussed in appendix B, delayed previews have an interesting theoretical link with graded comonads. The body of a lambda function may have a fully evaluated preview if it uses only variables that are bound by earlier let bindings, but it will typically be delayed. We consider a speculative design for an abstraction mechanism that better supports instant previews in appendix D.

### 5.2  Evaluation of previews

The evaluation of previews is defined in figure 8. Given a dependency graph $(V, E)$, the relation $v \Downarrow d$ evaluates a sub-expression corresponding to the node $v$ to a possibly delayed preview $d$. The nodes $V$ and edges $E$ of the graph are parameters of $\Downarrow$, but they do not change during the evaluation and so we do not explicitly write them.

The auxiliary relation $v \Downarrow_{\text{lift}} d$ always evaluates to a delayed preview. If the ordinary evaluation returns a delayed preview, so does the auxiliary relation (lift-expr). If the ordinary evaluation returns a value, the value is wrapped into a delayed preview requiring no variables (lift-prev). A node representing a value is evaluated to a value (val) and a node representing an unbound variable is reduced to a delayed preview that requires the variable and returns its value (var).

For member access, we distinguish two cases. If all arguments evaluate to values (member-val), then we use the evaluation relation defined by external libraries $\rightsquigarrow_\epsilon$ to immediately evaluate the member access and produce a value. If some of the arguments are delayed (member-expr), because the member access is inside the body of a lambda function, we produce a delayed member access expression that requires the union of the variables required by the individual arguments.

The evaluation of function values is similar, but requires three cases. If the body can be reduced to a value with no unbound variables (fun-val), we return a lambda function that returns the value. If the body requires only the bound variable (fun-bind), we return a lambda function with the delayed preview as the body. If the body requires further variables, the result is a delayed preview.

### 5.3 Caching of evaluated previews

For simplicity, the relation $\Downarrow$ in figure 8 does not specify how previews are cached. In practice, this is done by maintaining a lookup table from graph nodes $v$ to previews $p$. Whenever $\Downarrow$ is used to obtain a preview for a graph node, we first check the lookup table. If the preview has not been previously evaluated, we evaluate it and add it to the lookup. Cached previews can be reused in two ways. First, if the same sub-expression appears multiple times in the program, it will share a graph node and the preview will be resued. Second, when binding modified source code, the process reuses graph nodes and so previews are also reused during code editing.

## 6   Evaluating live programming environment

Computing previews using a dependency graph implements a correct and efficient optimization. In this section we show that this is the case, first theoretically in section 6.2, and then empirically in section 6.3. We also describe a case study where we developed an online service for data exploration based on the methods discussed in this paper (section 6.4).

### 6.1 Correctness of previews

To show that the previews are correct, we prove two properties. Correctness (theorem 6) guarantees that, the previews we calculate using a dependency graph are the same as the values we would obtain by evaluating the program directly. Determinacy (theorem 7) guarantees that previews assigned to a graph node based on an earlier graph are the same as previews that we would obtain afres using an updated graph.

To simplify the proofs, we consider programs without let bindings. Eliminating let bindings does not change the result of evaluation, as shown in lemma 2, and it also does not change the constructed dependency graph as shown below in lemma 3.

**Lemma 3** (Let elimintion for a dependency graph). *Given programs $p_1, p_2$ such that $p_1 \leadsto_{\mathsf{let}} p_2$ and a lookup table $\Delta_0$ then if $v_1; \ldots; v_n, (V, E) = \mathsf{bind\text{-}prog}_{\emptyset, \Delta_0}(p_1)$ and $v_1'; \ldots; v_n', (V', E') = \mathsf{bind\text{-}prog}_{\emptyset, \Delta_1}(p_2)$ such that $\Delta_1 = \mathsf{update}_{V,E}(\Delta_0)$ then for all $i$, $v_i = v_i'$ and also $(V, E) = (V', E')$.*

*Proof.* By analysis of the binding process. See appendix A.3. $\qquad\square$

The lemma 3 provides a way of removing let bindings from a program, such that the resulting dependency graph remains the same. Here, we bind the original program first, which adds the node for $e$ to $\Delta$. In our implementation, this is not needed because $\Delta$ is updated while the graph is being constructed using bind-expr. To keep the formalisation simpler, we separate the process of building the dependency graph and updating $\Delta$ and thus lemma 3 requires an extra binding step.

Now, we can show that, given a let-free expression, the preview obtained using a correctly constructed dependency graph is the same as the one we would obtain by directly evaluating the expression. This requires a simple auxiliary lemma.

**Lemma 4** (Lookup inversion). *Given $\Delta$ obtained using* update *in figure 7 then:*
- *If $v = \Delta(\mathsf{fun}(x), [(v_0, l_0)])$ then $v = \mathsf{fun}(x, s)$ for some $s$.*
- *If $v = \Delta(\mathsf{mem}(m), [(v_0, l_0), \ldots, (v_n, l_n)])$ then $v = \mathsf{mem}(m, s)$ for some $s$.*

*Proof.* By construction of $\Delta$ in figure 7. □

**Theorem 5** (Term preview correctness). *Given a term $t$ that has no free variables, together with a lookup table $\Delta$ obtained from any sequence of programs using* bind-prog *(figure 6) and* update *(figure 7), then let $v, (V, E) = \mathsf{bind\text{-}expr}_{\emptyset,\Delta}(t)$.*
*If $v \Downarrow p$ over a graph $(V, E)$ then $p = o$ for some value $o$ and $t \leadsto^* o$.*

*Proof.* By induction over the binding process. See appendix A.4. □

**Theorem 6** (Program preview correctness). *Consider a program $p = c_1; \ldots; c_n$ that has no free variables, together with a lookup table $\Delta_0$ obtained from any sequence of programs using* bind-prog *(figure 6) and* update *(figure 7). Assume a let-free program $p' = t_1; \ldots; t_n$ such that $p \leadsto^*_{\mathsf{let}} p'$.*
*Let $v_1; \ldots; v_n, (V, E) = \mathsf{bind\text{-}prog}_{\emptyset,\Delta_0}(p)$ and define updated lookup table $\Delta_1 = \mathsf{update}_{V,E}(\Delta_0)$ and let $v'_1; \ldots; v'_n, (V', E') = \mathsf{bind\text{-}prog}_{\emptyset,\Delta_1}(p)$.*
*If $v'_i \Downarrow p_i$ over a graph $(V', E')$ then $p_i = o_i$ for some value $o_i$ and $t_i \leadsto o_i$.*

*Proof.* Direct consequence of lemma 3 and theorem 5. □

Our implementation updates $\Delta$ during the recursive binding process and so a stronger version of the property holds: previews calculated over a graph obtained directly for the original program $p$ are the same as the values of the fully evaluated program. Our formalisation omits this for simplicity.

The second important property is determinacy, which makes it possible to cache the previews evaluated via $\Downarrow$ using the corresponding graph node as a lookup key.

**Theorem 7** (Preview determinacy). *For some $\Delta$ and for any programs $p, p'$, assume that the first program is bound, i.e. $v_1; \ldots; v_n, (V, E) = \mathsf{bind\text{-}prog}_{\emptyset,\Delta}(p)$, the graph node cache is updated $\Delta' = \mathsf{update}_{V,E}(\Delta)$ and the second program is bound, i.e. $v'_1; \ldots; v'_m, (V', E') = \mathsf{bind\text{-}prog}_{\emptyset,\Delta'}(p')$. Now, for any $v$, if $v \Downarrow p$ over $(V, E)$ then also $v \Downarrow p$ over $(V', E')$.*

*Proof.* By induction over $\Downarrow$, show that the same evaluation rules also apply over $(V', E')$. This is the case, because graph nodes added to $\Delta'$ by $\mathsf{update}_{V,E}$ are added as new nodes in $\mathsf{bind\text{-}prog}_{\emptyset,\Delta'}$ and nodes and edges of $(V, E)$ are unaffected. □

Edit contexts of expressions

$$K_e[-] \;=\; K_e[-].m(e_1,\ldots,e_n) \;\mid\; e.m(e_1,\ldots,e_{l-1},K_e[-],e_{l+1},\ldots,e_n) \;\mid\; -$$
$$K_c[-] \;=\; \mathsf{let}\, x = K_e[-] \;\mid\; K_e[-]$$

Code edit operations preserving preview for a sub-expression

(let-intro-var)  $\overline{c_1}; \ll e \gg; \overline{c_2}$ changes to $\overline{c_1}; \mathsf{let}\, x = e; \ll x \gg; \overline{c_2}$ where $x$ is fresh.

(let-intro-ins)  $\overline{c_1}; \overline{c_2}; \ll K_c[e] \gg; \overline{c_3}$ is changed to $\overline{c_1}; \mathsf{let}\, x = e; \overline{c_2}; \ll K_c[x] \gg; \overline{c_3}$ via a semantically non-equivalent expression $\overline{c_1}; \overline{c_2}; K_c[x]; \overline{c_3}$ where $x$ is free.

(let-intro-del)  $\overline{c_1}; \overline{c_2}; \ll K_c[e] \gg; \overline{c_3}$ is changed to $\overline{c_1}; \mathsf{let}\, x = e; \overline{c_2}; \ll K_c[x] \gg; \overline{c_3}$ via an expression $\overline{c_1}; \mathsf{let}\, x = e; \overline{c_2}; K_c[e]; \overline{c_3}$ with unused variable $x$.

(let-elim-del)  $\overline{c_1}; \mathsf{let}\, x = e; \overline{c_2}; \ll K_c[x] \gg; \overline{c_3}$ is changed to $\overline{c_1}; \overline{c_2}; \ll K_c[e] \gg; \overline{c_3}$ via a semantically non-equivalent expression $\overline{c_1}; \overline{c_2}; K_c[x]; \overline{c_3}$ where $x$ is free.

(let-elim-ins)  $\overline{c_1}; \mathsf{let}\, x = e; \overline{c_2}; \ll K_c[x] \gg; \overline{c_3}$ is changed to $\overline{c_1}; \overline{c_2}; \ll K_c[e] \gg; \overline{c_3}$ via an expression $\overline{c_1}; \mathsf{let}\, x = e; \overline{c_2}; K_c[e]; \overline{c_3}$ with unused variable $x$.

(edit-mem)  $\overline{c_1}; K_c[\ll e_0 \gg.m(\overline{e})]; \overline{c_2}$ is changed to $\overline{c_1}; K_c[\ll e_0 \gg.m'(\overline{e'})]; \overline{c_2}$

(edit-let)  $\overline{c_1}; \mathsf{let}\, x = e_1; \overline{c_2}; K_c[\ll e_2 \gg]; \overline{c_3}$ is changed to $\overline{c_1}; \mathsf{let}\, x = e'_1; \overline{c_2}; K_c[\ll e_2 \gg]; \overline{c_3}$     when $x \notin FV(e_2)$.

■ **Figure 9**  Code edit operations that preserve previously evaluated preview

The cache of previews (section 5.3) associates a preview $d$ with a node $v$ as the key. Theorem 7 guarantees that this is valid. As we update dependency graph during code editing, previous nodes will continue representing the same sub-expressions.

### 6.2 Reuse of previews

In this section, we identify a number of code edit operations where the previously evaluated values for a sub-expression can be reused. This includes the motivating example from section 1 where the data analyst extracted a constant into a let binding and modified a parameter of the last method call in a call chain.

The list of preview-preserving edits is shown in figure 9. It includes several ways of introducing and eliminating let bindings and edits where the analyst modifies an unrelated part of the program. The list is not exhaustive. Rather, it illustrates typical edits that the data analyst might perform when writing code. To express the operations we define an editing context $K$ which is similar to evaluation context $C$ from figure 3, but allows sub-expressions appearing anywhere in the program.

We use the notation $\ll e \gg$ to mark parts of expressions that are not recomputed during the edit; we write $\overline{c}$ and $\overline{e}$ for a list of commands and expressions, respectively. In some of the edit operations, we also specify an intermediate program that may be semantically different and only has a partial preview. This illustrates a typical way of working with code in a text editor using cut and paste. For example, in (let-intro-ins), the analyst cuts a sub-expression $e$, replaces it with a variable $x$ and then adds a let binding for a variable $x$ and inserts the expression $e$ from the clipboard. The (let-intro-del) operation captures the same edit, but done in a different order.

Theorem 8 proves that the operations given in figure 9 preserve the preview for a marked sub-expression. It relies on a lemma 13 given in appendix A.5 that generally characterizes one common kind of edits. Given two versions of a program that both contain the same sub-expression $e$, if the let bindings that define the values of variables used in $e$ do not change, then the graph node assigned to $e$ will be the same when binding the original and the updated program.

**Theorem 8** (Preview reuse). *Given the sequence of expressions as specified in figure 9, if the expressions are bound in sequence and graph node cache updated as specified in figure 7, then the graph nodes assigned to the specified sub-expressions are the same.*

*Proof.* Cases (edit-let) and (edit-mem) are direct consequences of lemma 13; for (let-intro-var), the node assigned to $x$ is the node assigned to $e$ which is the same as before the edit from lemma 13. Cases (let-intro-ins) and (let-intro-del) are similar to (let-intro-var), but also require using induction over the binding of $K_c[e]$. Finally, cases (let-elim-ins) and (let-elim-del) are similar and also use lemma 13 together with induction over the binding of $K_c[x]$. □

### 6.3 Empirical evaluation of efficiency

The key performance claim about our method of providing instant feedback is that it is more efficient than recomputing values for the whole program (or the current command) after every keystroke. In the previous section, we formally proved that this is true and gave examples of code edit operations that do not cause recomputation. In this section, we further support this claim with an empirical evaluation. The purpose of this section is not to precisely evaluate overheads of our implementation, but to compare how much recomputation different evaluation strategies perform.

For the purpose of the evaluation, we use a simple image manipulation library that provides operations for loading, greyscaling, blurring and combining images. We compare delays in updating the preview for three different evaluation strategies, while performing the same sequence of code edit operations. Using image processing as an example gives us a way to visualize the reuse of previously computed values. As in a typical data exploration scenario, the individual operations are relatively expensive compared to the overheads of building the dependency graph.

To avoid distractions when visualizing the performance, we update the preview after complete tokens are added rather than after individual keystrokes. Figure 10 shows the sequence of edits that we use to measure the delays in updating a instant preview. We first enter an expression to load, greyscale and blur an image (1) then introduce let binding (2) and add more operations (3). Finally, we extract one of the parameters into a variable (4). Most of the operations are simply adding code, but there are two cases where we modify existing code and change value of a parameter for blur and combine immediately after (1) and (3), respectively.

(1) Enter the following code and then change parameter of blur from 4 to 8:

```
image.load("shadow.png").greyScale().blur(4)
```

(2) Assign the result to a variable and start writing code for further operations:

```
let shadow = image.load("shadow.png").greyScale().blur(8)
shadow.combine
```

(3) Finish code to combine two images and change parameter of combine from 20 to 80:

```
let shadow = image.load("shadow.png").greyScale().blur(8)
shadow.combine(image.load("pope.png"), 20)
```

(4) Extract the parameter of combine to a let bound variable:

```
let ratio = 80
let shadow = image.load("shadow.png").greyScale().blur(8)
shadow.combine(image.load("pope.png"), ratio)
```

■ **Figure 10**   Code edit operations that are used in the experimental evaluation

We implement the algorithm described in section 4 section 5 in a simple web-based environment that allows the user to modify code and explicitly trigger recomputation. It then measures time needed to recompute values for the whole program and displays the resulting image. If the parsing fails, we record only the time taken by parsing attempt. We compare the delays of three different evaluation strategies:

**Call-by-value**  Following the semantics in section 3.2, all sub-expressions are evaluated before an expression. This is often wasteful. For example, we parse the expression image.load("shadow.png").blur as a member access with no arguments. The evaluation loads the image, but then fails because blur requires one argument.

**Lazy**  To address the wastefulness of call-by-value strategy, we simulate lazy evaluation by implementing a version of the image processing library where operations build a delayed computation and only evaluate it when rendering an image. Using this strategy, failing computations do not perform unnecessary work.

**Live**  Finally, we use the algorithm described in section 4 section 5. The cache is empty at the beginning of the experiment and we update it after each token is added. This is the only strategy where evaluation does not start afresh after reparsing code.

The experimental environment is implemented in F# and compiled to JavaScript using the Fable compiler. We run the experiments in Firefox (version 64.0.2, 32 bit)) on Windows 10 (build 1809, 64 bit) on Intel Core i7-7660U CPU with 16 GB RAM.

Figure 11 shows times needed to recompute previews after individual tokens are added, deleted or modified, according to the script in figure 10, resulting in 38 measurements. We mark a number of notable points in the chart:

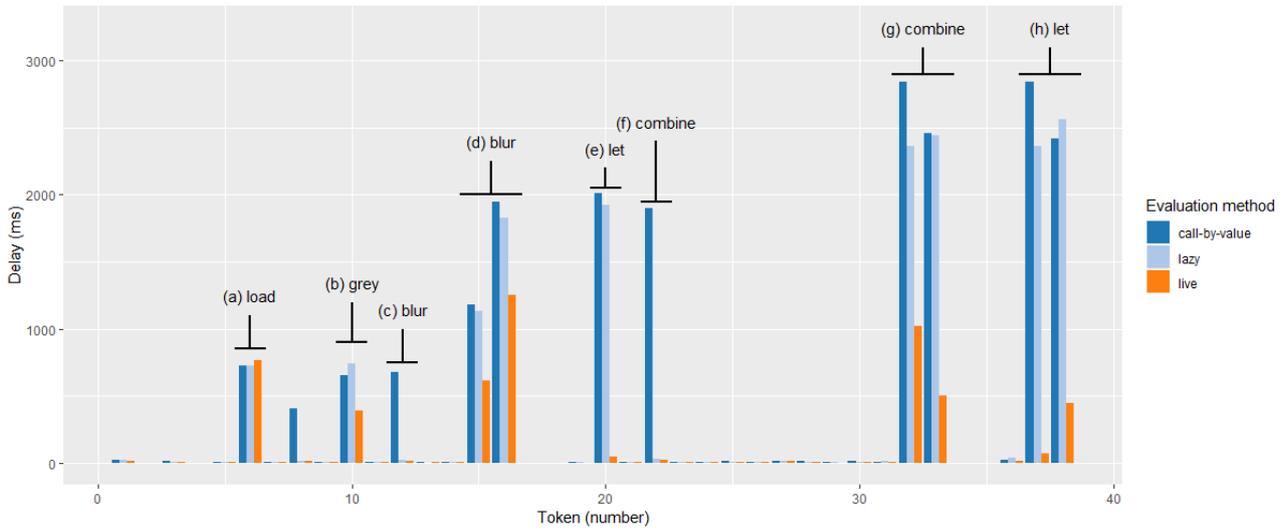

■ **Figure 11** Time required to recompute the results of a sample program after individual tokens are added or modified for three different evaluations strategies.

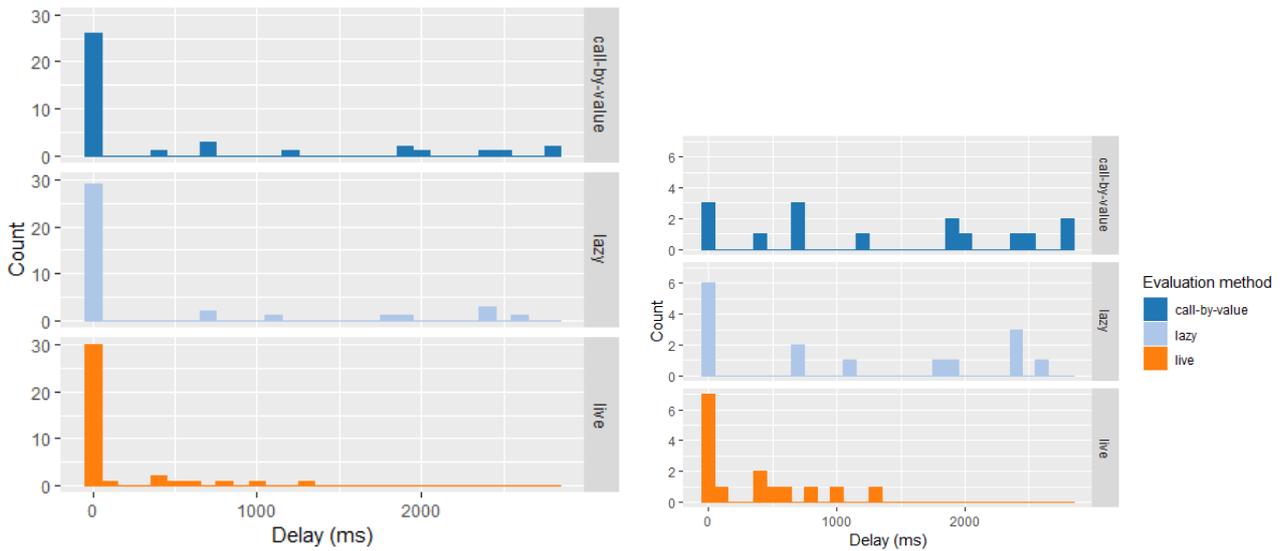

■ **Figure 12** Distribution of delays incurred when updating previews. We show a histogram computed from all delays (left) and only from delays larger than 15 ms (right).

a. Loading image for the first time incurs small extra overhead in the live strategy.

b. Greyscaling using the live strategy does not need to re-load the image.

c. Accessing the blur member without arguments causes delay for call-by-value.

d. When varying the parameter of blur, the live strategy reuses the greyscaled image.

e. Introducing let binding does not cause recomputation when using live strategy.

f. As in (c), accessing a member without an argument only affects call-by-value.

g. The live strategy is much faster when varying the parameter of combine.

h. Introducing let binding, again, causes full recomputation for lazy and call-by-value.

A summarized view of the delays is provided in figure 12, which shows histograms illustrating the distribution of delays for each of the three evaluation methods. A large proportion of delays is very small (less than 15 ms) because the parser used in our experimental environment often fails (e.g. for unclosed parentheses). The histogram on the right summarizes only delays for edit operations where the delay for at least one of the strategies was over 15 ms. The histogram shows that the live strategy eliminates the longest delays (by caching partial results), with the exception of a few where the underlying operation takes a long time (such as blurring the image). The results would be even more significant with an error-recovering parser.

The purpose of our experiment is not to exactly assess the overhead of our implementation. Our goal is to illustrate how often can previously evaluated results be reused and the impact this has when writing code. The experiment presented in this section is small-scale, but it is sufficient for this purpose. When recomputing results after every edit using the *call-by-value* strategy, the time needed to update results grows continually. The *lazy* strategy removes the overhead for programs that fail, but keeps the same trend. Our *live* strategy reuses values computed previously. Consequently, expensive operations such as (d) and (g) in figure 11 are significantly faster, because they do not need to recompute operations done previously when writing the code. As shown in figure 12, there are almost no very expensive operations (taking over 1 second) in the *live* strategy in contrast to several in the other two strategies.

### 6.4 Transparent tools for data journalism

In section 2.1, we motivated our work by considering how journalists explore open data. In addition to the theoretical and experimental work presented in this paper, we also implemented an online data exploration environment, equipped with live editor for The Gamma language that provides instant feedback during coding. The environment uses the principles presented in this paper to build a more comprehensive system that allows users, such as journalists, to analyse, summarize and visualize open data. In this section, we briefly report on our experience with the system. Two screenshots shown in figure 13 illustrate a number of interesting features:

- The left screenshot uses a type provider for data aggregation [45]. Type provider are treated as an external library (with objects, members and reduction relation). Type providers also rely on type information to provide editor auto-complete, which we support by implementing type checking over a dependency graph (appendix C).

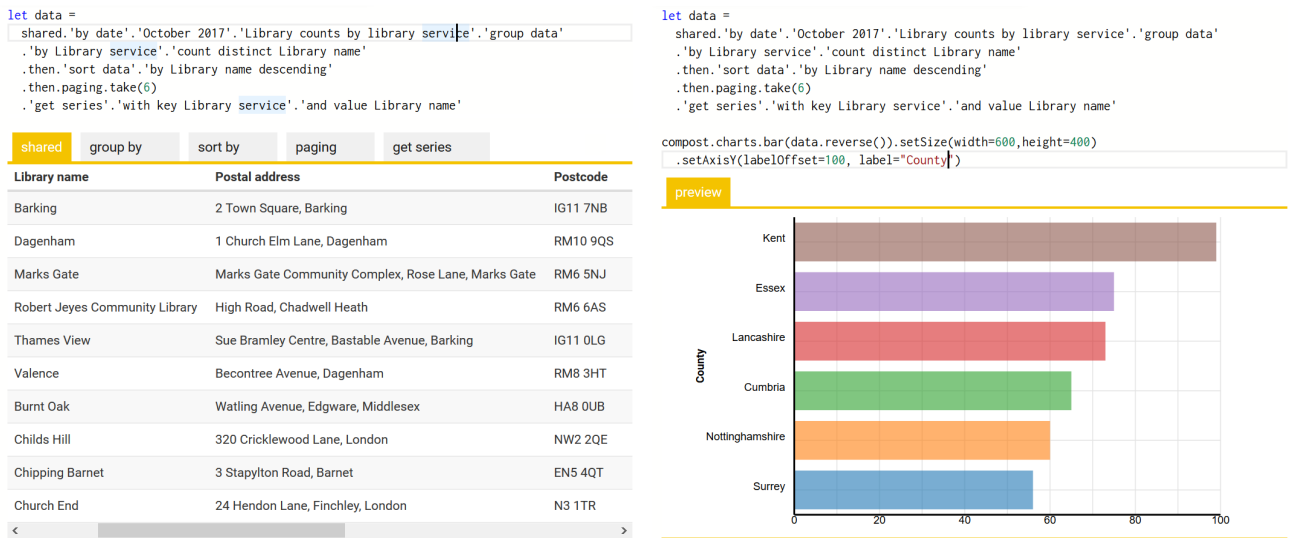

■ **Figure 13** Data analysis counting the number of libraries per county in the UK. We load and aggregate data (left) and then create a chart with a label (right). The example has been created using our tool discussed in section 6.4 by a non-programmer and is available at http://gallery.thegamma.net/73/.

- When displaying instant preview for code written using the data aggregation type provider (left), our environment generates tabes that show individual steps of the data transformation. A tab is selected based on the cursor position and shows a preview for the current sub-expression (e.g. raw data before grouping).
- Another external library provides support for charting (right). Here, the environment displays the result of evaluating the whole command. The screenshot shows a case where the user modifies parameters of the chart. Thanks to our live evaluation strategy, this is done efficiently without reevaluating the data transformation.

The environment is available at http://gallery.thegamma.net. A user study to evaluate the usability of the system from a human-computer interaction perspective is left for future work. As an anecdotal evidence, the code in figure 13 was developed by an attendee of a Mozilla Festival 2017 who had no prior programming experience.

## 7  Related and future work

Simple data exploration performed, for example, by journalists [20] is done either programmatically or using spreadsheets. The latter is easy but error-prone while the former requires expert programming skills. We aim to bring liveness of spreadsheets to programmatic data exploration. This requires extending recomputation as done in spreadsheets [52] to code written in an ordinary text editor.

**Notebooks and data science tools.**   Visual data exploration tools are interactive [11, 24, 60] and some can export the workflow as a script [26], but data analysts who prefer code typically resort to notebook systems such as Jupyter or R Markdown [5, 29]. Those are text-based, but have a limited model of recomputation. Users structure code in cells and manually reevaluate cells. Many notebook systems are based on the REPL (read-eval-print-loop) [16, 34] and do not track dependencies between cells, which can lead to well-documented inconsistencies [30, 46, 49].

Ideas such as dependency tracking and efficient recomputation exist in visual data exploration tools [11, 24, 60] and scientific workflow systems [6, 41]. The implementation techniques are related to our work, but we focus on text-based scripts. Tempe [13] focuses on streaming data, but provides a text-based scripting environment with automatic code update; its usability in contrast to REPLs has been empirically evaluated [12].

**Live and exploratory programming.** Live programming based on textual programs has been popularised by Victor [57, 58] and is actively developed in domains such as live coded music [1, 51]. The notions of exploratory and live programming have been extensively studied [50]. The notion of exploratory programming has recently been analysed from the perspective of human-computer interaction [28], which led to new tools [27], complementary to our instant previews. Kubelka, Robbes, and Bergel [31] review the use of live programming in a Smalltalk derived environment. Lighttable [19] and Chrome DevTools provide limited instant previews akin to those presented in this paper, but without well specified recomputation model. Finally, work on keeping state during code edits [8, 35] would be relevant for supporting streaming data.

A more principled approach can be used by systems based on structured editing [33, 42, 44, 55] where code is only modified via known operations with known effect on the computation graph (e.g. "extract variable" has no effect on the result; "change constant value" forces recomputation of subsequent code). This can be elegantly implemented using bi-directional lambda calculus [43], but it also makes us consider more human-centric abstractions [14, 15] further discussed in appendix D.

**Incremental computation and dependency analysis.** Work on self-adjusting and incremental computation [2, 23] handles recomputation when the program stays the same, but input changes. Most incremental systems, e.g. [3, 4, 22] evaluate the program and use programmer-supplied information to build a dependency graph, whereas our system uses static code analysis;[23] implement a small adaptive interpreter that treats code as changing input data, suggesting an implementation technique for live programming systems. Our use of dependency graphs [32] is static and first-order and can be seen as a form of program slicing [59], although our binding process is more directly inspired by Roslyn [40], which uses it for efficient background type-checking.

**Semantics and partial evaluation.** The evaluation of previews is a form of partial evaluation [10], done in a way that allows caching. This can be done implicitly or explicitly in the form of multi-stage programming [56]. Semantically, the evaluation of previews can be seen as a modality and delayed previews are linked to contextual modal type theory [39], formally modelled using comonads [17].

## 8  Summary

One of the aspects that make spreadsheets easier to use than programming tools is that they provide instant feedback. We aim to make programming tools as instant as spreadsheets. We described a number of key aspects of simple data analyses such as those done by journalists and then used our observations to build both theory and simple practical data analytics tools.

Our *data exploration calculus* is a simple formally tractable language for data exploration. The calculus captures key observations about simple data analyses. They rely on logic defined by external libraries, implement few abstractions and are written as lists of commands.

Our main technical contribution is a *instant preview* mechanism that efficiently evaluates code during editing and instantly provides a preview of the result. We allow users to edit code in unconstrained way in an text editor, which makes this particularly challenging. The key trick is to separate the process into a fast *binding phase*, which constructs a dependency graph and a slower *evaluation phase* that can cache results. This makes it possible to quickly parse updated code, reconstruct dependency graph and compute preview using previous, partially evaluated, results.

We evaluated our approach in three ways. First, we proved that our mechanism is correct and that it reuses evaluated values for many common code edit operations. Second, we conducted an experimental study that illustrates how often are previously evaluated results reused during typical programming scenario. Thirdly, we used our research as a basis for online data exploration environment, which shows the practical usability of our work.

**Acknowledgements**   We thank Dominic Orchard, Stephen Kell, Roly Perera and Jonathan Edwards for many discussions about the work presented in the paper. Dominic Orchard provided invaluable feedback on earlier version of the paper. The paper also benefited from suggestions made by anonymous reviewers of The Programming Journal as well as reviwers of earlier versions of the paper. This work was partly supported by The Alan Turing Institute under the EPSRC grant EP/N510129/1.

## A    Details of proofs

### A.1    Normalization for data exploration calculus

**Theorem 9** (Normalization). *For all $p$, there exists $n$ and $o_1, \ldots, o_n$ such that $p \leadsto^* o_1; \ldots; o_n$.*

*Proof.* We define size of a program in data exploration calculus as follows:

$$
\begin{array}{rcl}
\mathsf{size}(c_1; \ldots; c_n) & = & 1 + \Sigma_{i=1}^{n}\mathsf{size}(c_i) \\
\mathsf{size}(\mathsf{let}\ x = t) & = & 1 + \mathsf{size}(t) \\
\mathsf{size}(e_0.m(e_1, \ldots, e_n)) & = & 1 + \Sigma_{i=0}^{n}\mathsf{size}(e_i) \\
\mathsf{size}(\lambda x \to e) & = & 1 + \mathsf{size}(e) \\
\mathsf{size}(o) = \mathsf{size}(x) & = & 1
\end{array}
\tag{10}
$$

The property holds because, first, both **(let)** and **(external)** decrease the size of the program and, second, a program is either fully evaluated, i.e. $o_1; \ldots; o_n$ for some $n$ or, it can be reduced using one of the reduction rules. □

### A.2    Let elimmination for a program

**Lemma 10** (Let elimination for a program). *Given any program $p$ such that $p \leadsto^* o_1; \ldots; o_n$ for some $n$ and $o_1, \ldots, o_n$ then if $p \leadsto_{\mathsf{let}} p'$ for some $p'$ then also $p' \leadsto^* o_1; \ldots; o_n$.*

*Proof.* The elimination of let binding transforms a program $c_1; \ldots; c_{k-1};$ $\mathsf{let}\ x = t;\ c_{k+1}; \ldots; c_n$ to a program $c_1; \ldots; c_{k-1};\ t; c_{k+1}[x \leftarrow t]; \ldots; c_n[x \leftarrow t]$. The reduction steps for the new program can be constructed using the steps of $p \leadsto^* o_1; \ldots; o_n$. The new command $t$ reduces to an object $o$ using the same steps as the original term $t$ in $\mathsf{let}\ x = t$ but with context $C_c = -$ rather than $C_c = \mathsf{let}\ x = -$; the terms $t$ introduced by substitution also reduce using the same steps as before, but using contexts in which the variable $x$ originally appeared. □

### A.3    Let elimination for a dependency graph

**Lemma 11** (Let elimintion for a dependency graph). *Given programs $p_1, p_2$ such that $p_1 \leadsto_{\mathsf{let}} p_2$ and a lookup table $\Delta_0$ then if $v_1; \ldots; v_n, (V, E) = \mathsf{bind\text{-}prog}_{\emptyset, \Delta_0}(p_1)$ and $v_1'; \ldots; v_n', (V', E') = \mathsf{bind\text{-}prog}_{\emptyset, \Delta_1}(p_2)$ such that $\Delta_1 = \mathsf{update}_{V,E}(\Delta_0)$ then for all $i$, $v_i = v_i'$ and also $(V, E) = (V', E')$.*

*Proof.* Assume $p_1 = c_1; \ldots; c_{k-1}; \mathsf{let}\ x = e; c_{k+1}; \ldots; c_n$ and the let binding is eliminated resulting in $p_2 = c_1; \ldots; c_{k-1}; e; c_{k+1}[x \leftarrow e]; \ldots; c_n[x \leftarrow e]$. When binding $p_1$, the case $\mathsf{bind\text{-}prog}_{\Gamma, \Delta}(\mathsf{let}\ x = e)$ is handled using (7) and the node resulting from binding $e$ is added to the graph $V, E$. It is then referenced each time $x$ appears in subsequent commands $c_{k+1}; \ldots; c_n$. When binding $p_2$, the node resulting from binding $e$ is a primitive value or a node already present in $\Delta_1$ (added by $\mathsf{update}_{V,E}$) and is reused each time $\mathsf{bind\text{-}expr}_{\Gamma, \Delta_1}(e)$ is called. □

### A.4 Term preview correctness

**Theorem 12** (Term preview correctness). *Given a term $t$ that has no free variables, together with a lookup table $\Delta$ obtained from any sequence of programs using* bind-prog *(figure 6) and* update *(figure 7), then let* $\nu, (V,E) =$ bind-expr$_{\emptyset,\Delta}(t)$. *If* $\nu \Downarrow p$ *over a graph* $(V,E)$ *then* $p = o$ *for some value* $o$ *and* $t \rightsquigarrow^* o$.

*Proof.* When combining recursively constructed sub-graphs, the bind-expr function adds new nodes and edges leading from those new nodes. Therefore, an evaluation using $\Downarrow$ over a sub-graph will also be valid over the new graph – the newly added nodes and edges do not introduce non-determinism to the rules given in figure 8.

We prove a more general property showing that for any $e$, its binding $\nu, (V,E) =$ bind-expr$_{\emptyset,\Delta}(e)$ and any evaluation context $C$ such that $C[e] \rightsquigarrow o$ for some $o$, one of the following holds:

a. If $FV(e) = \emptyset$ then $\nu \Downarrow p$ for some $p$ and $C[p] \rightsquigarrow o$

b. If $FV(e) \neq \emptyset$ then $\nu \Downarrow [\![e_p]\!]_{FV(e)}$ for some $e_p$ and $C[e_p] \rightsquigarrow o$

In the first case, $p$ is a value, but it is not always the case that $e \rightsquigarrow^* p$, because $p$ may be lambda function and preview evaluation may reduce sub-expression in the body of the function. Using a context $C$ in which the value reduces to an object avoids this problem.

The proof of the theorem follows from the more general property. Using a context $C[-] = -$, the term $t$ reduces $t \rightsquigarrow^* t' \rightsquigarrow_\epsilon o$ for some $o$ and the preview $p$ is a value $o$ because $C[p] = p = o$. The proof is by induction over the binding process, which follows the structure of the expression:

(1) bind-expr$_{\Gamma,\Delta}(e_0.m(e_1, \ldots, e_n))$ – Here $e = e_0.m(e_1, \ldots, e_n)$, $\nu_i$ are graph nodes obtained by induction for expressions $e_i$ and $\{(\nu, \nu_0, \mathsf{arg}(0)), \ldots, (\nu, \nu_n, \mathsf{arg}(n))\} \subseteq E$. From lookup inversion lemma 4, $\nu = \mathsf{mem}(m, s)$ for some $s$.

  If $FV(e) = \emptyset$, then $\nu_i \Downarrow p_i$ for $i \in 0 \ldots n$ and $\nu \Downarrow p$ using (mem-val) such that $p_0.m(p_1, \ldots, p_n) \rightsquigarrow p$. From induction hypothesis and *compositionality* of external libraries (definition 2), it holds that for any $C$ such that $C[e_0.m(e_1, \ldots, e_n)] \rightsquigarrow o$ for some $o$ then also $C[p_0.m(p_1, \ldots, p_n)] \rightsquigarrow C[p] \rightsquigarrow o$.

  If $FV(e) \neq \emptyset$, then $\nu_i \Downarrow_{\mathsf{lift}} [\![e_i']\!]$ for $i \in 0 \ldots n$ and $\nu \Downarrow [\![e_0'.m(e_1', \ldots, e_n')]\!]_{FV(e)}$ using (mem-expr). From induction hypothesis and *compositionality* of external libraries (definition 2), it holds that for any $C$ such that $C[e_0.m(e_1, \ldots, e_n)] \rightsquigarrow o$ for some $o$ then also $C[e_0'.m(e_1', \ldots, e_n')] \rightsquigarrow o$.

(2) bind-expr$_{\Gamma,\Delta}(e_0.m(e_1, \ldots, e_n))$ – This case is similar to (1), except that the fact that $\nu = \mathsf{mem}(m, s)$ holds by construction, rather than using lemma 4.

(3) bind-expr$_{\Gamma,\Delta}(\lambda x \to e_b)$ – Here $e = \lambda x \to e_b$, $\nu_b$ is the graph node obtained by induction for the expression $e_b$ and $(\nu, \nu_b, \mathsf{body}) \in E$. From lookup inversion lemma 4, $\nu = \mathsf{fun}(x, s)$ for some $s$. The evaluation can use one of three rules:

  i. If $FV(e) = \emptyset$ then $\nu_b \Downarrow p_b$ for some $p_b$ and $\nu \Downarrow \lambda x \to p_b$ using (fun-val). Let $e_b' = p_b$.

  ii. If $FV(e_b) = \{x\}$ then $\nu_b \Downarrow [\![e_b']\!]_x$ for some $e_b'$ and $\nu \Downarrow \lambda x \to e_b'$ using (fun-bind).

  iii. Otherwise, $\nu_b \Downarrow [\![e_b']\!]_{x,\Gamma}$ for some $e_b'$ and $\nu \Downarrow [\![\lambda x \to e_b']\!]_\Gamma$ using (fun-expr).

For i.) and ii.) we show that a.) is the case; for iii.) we show that b.) is the case; that is for any $C$, if $C[\lambda x \rightarrow e_b] \rightsquigarrow o$ then also $C[\lambda x \rightarrow e'_b] \rightsquigarrow o$. For a given $C$, let $C'[-] = C[\lambda x \rightarrow -]$ and use the induction hypothesis, i.e. if $C'[e_b] \rightsquigarrow o$ for some $o$ then also $C'[e'_b] \rightsquigarrow o$.

(4) bind-expr$_{\Gamma, \Delta}(\lambda x \rightarrow e)$ – This case is similar to (3), except that the fact that $v = $ fun$(x, s)$ holds by construction, rather than using lemma 4.

(5) bind-expr$_{\Gamma, \Delta}(o)$ – In this case $e = o$ and $v = $ val$(o)$ and val$(o) \Downarrow o$ using (val) and so the case a.) trivially holds.

(6) bind-expr$_{\Gamma, \Delta}(x)$ – The initial $\Gamma$ is empty, so $x$ must have been added to $\Gamma$ by case (3) or (4). Hence, $v = $ var$(x)$, $v \Downarrow [\![x]\!]_x$ using (var) and so $e_p = e = x$ and the case b.) trivially holds.

$\square$

### A.5 Binding sub-expressions

**Lemma 13** (Binding sub-expressions). *Assume we have programs $p_1, p_2$ such that $p_1 = c_1; \ldots; c_k; K_c[e]; c_{k+1}; \ldots; c_n$ and $p_2 = c'_1; \ldots; c'_k; K'_c[e]; c'_{k+1}; \ldots; c'_n$ and $I \subseteq \{1 \ldots k\}$ such that $\forall i \in I . c_i = c'_i$ and for each $x \in \bigcup_{i \in I} FV(c_i) \cup FV(e)$ there exists $j \in I$ such that $c_j = $ let $x = e$ for some $e$. Given any $\Delta$, assume that the the first program is bound, i.e. $v_1; \ldots; v_n, (V, E) = $ bind-prog$_{\emptyset, \Delta}(p_1)$, the cache is updated $\Delta' = $ update$_{V, E}(\Delta)$ and the second program is bound, i.e. $v'_1; \ldots; v'_n, (V', E') = $ bind-prog$_{\emptyset, \Delta'}(p_2)$.*

*Now, assume $v, G = $ bind-expr$_{\Gamma, \Delta}(e)$ and $v', G' = $ bind-expr$_{\Gamma', \Delta'}(e)$ are the recursive calls to bind $e$ during the first and the second binding, respectively. Then, the graph nodes assigned to the sub-expression $e$ are the same, i.e. $v = v'$.*

*Proof.* First, assuming that $\forall x \in FV(e) . \Gamma(x) = \Gamma'(x)$, we show by induction over the binding process of $e$ for the first program that the result is the same. In cases (1) and (3), the updated $\Delta'$ contains the required key and so the second binding proceeds using the same case. In cases (2) and (4), the second binding reuses the node created by the first binding using case (1) and (3), respectively. Cases (5) and (6) are the same.

Second, when binding let bindings in $c_1; \ldots; c_k$, the initial $\Gamma = \emptyset$ during both bindings. Nodes added to $\Gamma$ and $\Gamma'$ for commands $c_j$ such that $j \in I$ are the same and nodes added for remaining commands do not add any new nodes referenced from $e$ and so $v = v'$ using the above. $\square$

## B  Theories of delayed previews

The operational semantics presented in this paper serves two purposes. It gives a simple guide for implementing text-based live programming environments for data science and we use it to prove that our optimized way of producing instant previews is correct. However, some aspects of our mechanism are related to important work in semantics of programming languages and deserve to be mentioned.

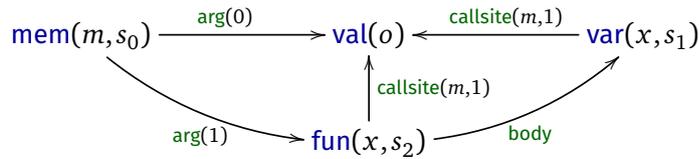

■ **Figure 14** Dependency graph for $o.m(\lambda x \to x)$ with a newly added callsite edges.

The construction of delayed previews is related to meta-programming. Assuming we have delayed previews $[\![e_0]\!]_x$ and $[\![e_1]\!]_y$ and we invoke a member $m$ on $e_0$ using $e_1$ as an argument. To do this, we construct a new delayed preview $[\![e_0.m(e_1)]\!]_{x,y}$. This operation is akin to expression splicing from meta-programming [53, 56].

The semantics of delayed previews can be more formally captured by Contextual Modal Type Theory (CMTT) [39] and comonads [17]. In CMTT, $[\Psi]A$ denotes that a proposition $A$ is valid in context $\Psi$, which is similar to our delayed previews written as $[\![A]\!]_\Psi$. CMTT defines rules for composing context-dependent propositions that would allow us to express the splicing operation used in (mem-expr). In categorical terms, the context-dependent proposition can be modelled as a graded comonad [18, 38]. The evaluation of a preview with no context dependencies (built implicitly into our evaluation rules) corresponds to the counit operation of a comonad and would be explicitly written as $[\![A]\!]_\emptyset \to A$.

## C    Type checking

Instant previews give analysts quick feedback when they write incorrect code, but having type information is still valuable. First, it can help give better error messages. Second, types can be used to provide auto-complete – when the user types '.' we can offer available members without having to wait until the value of the object is available.

**Revised dependency graph.**    Type checking of small programs is typically fast enough that no caching is necessary. However, The Gamma supports *type providers* [9, 54], which can generate types based on an external file or a REST service call, e.g. [47]. For this reason, type checking can be relatively time consuming and can benefit from the same caching facilities as those available for instant previews.

Adding type checking requires revising the way we construct the dependency graph introduced in section 4. Previously, a variable bound by a lambda function had no dependencies. However, the type of the variable depends on the context in which it appears. Given an expression $o.m(\lambda x \to x)$, we infer the type of $x$ from the type of the first argument of the member $m$. A variable node for $x$ thus needs to depend on the call site of $m$. We capture that by adding an edge callsite$(m, i)$ from $x$ to $o$ which indicates that $x$ is the input variable of a function passes as the $i^{\text{th}}$ argument to the $m$ member of the expression represented by the target node. We also add callsite$(m, i)$

$$(\text{var}) \ \frac{\begin{array}{c} (\text{var}(x,s), v, \text{callsite}(m,i)) \in E \\ v \vdash \{.., m : (\tau_1, \ldots, \tau_k) \to \tau, ..\} \qquad \tau_i = \tau' \to \tau'' \end{array}}{\text{var}(x,s) \vdash \tau'}$$

$$(\text{mem}) \ \frac{\begin{array}{c} \forall i \in \{0 \ldots k\}. (\text{mem}(m,s), v_i, \text{arg}(i)) \in E \\ v_0 \vdash \{.., m : (\tau_1, \ldots, \tau_k) \to \tau, ..\} \qquad v_i \vdash \tau_i \end{array}}{\text{mem}(m,s) \vdash \tau}$$

$$(\text{fun}) \ \frac{\begin{array}{c} \{(\text{fun}(x,s), v_b, \text{body}), (\text{var}(x,s), v_c, \text{callsite}(m,i))\} \subseteq E \\ v_c \vdash \{.., m : (\tau_1, \ldots, \tau_k) \to \tau, ..\} \qquad \tau_i = \tau' \to \tau'' \qquad v_b \vdash \tau'' \end{array}}{\text{fun}(x,s) \vdash \tau' \to \tau''}$$

■ **Figure 15**  Rules that define type checking of terms and expressions over a dependency graph $(V, E)$

as an edge from the node of the function. Figure 14 shows the revised dependency graph for $o.m(\lambda x \to x)$.

**Type checking.**  The structure of typing rules is similar to the evaluation relation $v \Downarrow d$ defined earlier. Given a dependency graph $(V, E)$, we define typing judgements in the form $v \vdash \tau$. The type $\tau$ can be a primitive type, a function $\tau \to \tau$ or an object type $\{m_1 : \sigma_1, \ldots, m_n : \sigma_n\}$ with member types $\sigma = (\tau_1, \ldots, \tau_n) \to \tau$.

The typing rules for variables, functions and member access are shown in figure 15. When type checking a member access (mem), we find its dependencies $v_i$ and check that the instance is an object with the required member $m$. The types of arguments of the member then need to match the types of the remaining (non-instance) nodes. Type checking a function (fun) and a variable (var) is similar. In both cases, we follow the callsite edge to find the member that accepts the function as an argument and use the type of the argument to check the type of the function or infer the type of the variable.

The results of type checking can be cached and reused in the same way as instant previews, although we leave out the details. A property akin to correctness (theorem 6) requires defining standard type checking over the structure of expressions, which we also omit for space reasons.

## D  Feedback-friendly abstraction

The data analysis by Financial Times in section 2.1 illustrates why notebook users often avoid abstraction. Wrapping code in a function makes it impossible to split code into cells and see results of intermediate steps. Instead, the analysis used a global variable with possible values in a comment.

Providing instant previews inside ordinary functions is problematic, because we do not have readily available values for input parameters and our mechanism for lambda functions only provides delayed previews inside body of a function. We believe that extending the data exploration calculus with an abstraction mechanism that would support development with instant feedback is an interesting design problem and we briefly outline a possible solution here.

Data scientists often write code interactively using a sample data set and, when it works well, wrap it into a function that they then call on other data. Similarly, spreadsheet users often write equation in the first row of a table and then use the "drag down" operation to apply it to other rows. One way of adding similar functionality to the data exploration calculus is to label a sequence of commands such that the sequence can be reused later with different inputs:

$$
\begin{aligned}
p &= c_1; \ldots; c_n \\
c &= \mathsf{let}\, x = t \mid t \mid lbl\colon p \mid lbl
\end{aligned}
$$

We introduce two new kinds of commands: a labelled sequence of commands and a reference to a label. When evaluating, the command $lbl$ is replaced with the associated sequence of commands $p$ before any other reductions. Consequently, variables used in the labelled block are dynamically scoped and we can use let binding to redefine a value of a variable before invoking the block repeatedly. The correct use of dynamic scoping can be checked using coeffects [48].

This minimalistic abstraction mechanism supports code reuse without affecting how instant previews are computed. Commands in a labelled block require variables to be defined before the block. Those define sample data for development and can be redefined before reusing the block. We intend to implement this mechanism in a future version of our data exploration environment (section 6.4).

## About the author

**Tomas Petricek** is a lecturer at University of Kent. He is working on making programming and data science easier and more accessible. His research interests span a range of areas including theory of programming languages, tools for data science as well as history and philosophy of programming. Previously, he contributed to functional programming and the development of the F# language and type providers at Microsoft Research and obtained PhD from University of Cambridge for his work on coeffects, a theory of context-aware programming languages. Along the way, he became interested in understanding programming through the perspective of history and philosophy of science and wrote papers about the evolution of programming concepts such as types and errors. Contact him at tomas@tomasp.net.